# Additional Quantum Numbers for Electron Orbitals of Nanoparticles

V.G.Yarzhemsky

Kurnakov Institute of General and Inorganic Chemistry of RAS, 31 Leninsky, 119991 Moscow, Russia, e-mail vgyar@igic.ras.ru

Abstract

It is shown that in large symmetrical nanoparticles additional quantum numbers are required to label unambiguously symmetry adapted linear combinations of the wave functions. It is obtained that the labels of irreducible representations (IRs) of intermediate subgroups (between the whole symmetry group and local subgroup) can be used for complete classification of states in the case of repeating IRs. The examples of $O_h$ and $I_h$ groups are considered and the connection of additional quantum numbers with local interactions is discussed.

PACS 02.20.Bd, 61.46-w, 71.10-m

KEY WORDS: nanoparticles, electron orbitals, group theory, induced representations

Introduction

The symmetry of nanostructures, as well as the symmetry of molecules is described by point group symmetry, and there is a large variety of structures described by the same symmetry group, but characterizing by different local symmetry subgroups. For example, in recently discovered onion like structure $[AsNi_{12}As_{20}]^{3-}$ [1] the local symmetry of the central atom is $I_h$, the symmetry of each atom in $Ni_{12}$ cage is $C_{5v}$ and the local symmetry of atoms in $As_{20}$ cage is $C_{3v}$. Icosahedral gold nanoparticles $Au_{32}$, $Au_{42}$ and $Au_{72}$ [2] consist of several different orbits of gold atoms. Note that orbit is the set of points, transforming into each other under the action of group elements [3]. Direct analysis shows that in some cases repeating IRs (irreducible representations) appear in symmetry adapted linear combinations (SALC) constructed from the atomic wavefunctions of one same orbit [4]. Intermediate continuous [5] and finite [6] groups use usually used to label repeating IRs in the reduction from a larger symmetry group to the symmetry group of physical Hamiltonian. Intermediate groups can be also used for labeling of repeating IRs in the induction from a subgroup to the whole group [7].The induced representation method is a powerful technique for the analysis of the wavefunctions of molecules [7,8] and other multicentre systems with physically distinguished subgroups of symmetry [9-11]. This method was further developed and applied to labeling of states in molecules in the works of Newman et.al. [12-14].

In the present work the induced representation method with intermediate subgroup is applied for labeling of repeating IRs in SALCs of symmetrical nanostructures and groups $O_h$ and $I_h$ are taken as examples. It is shown that in many cases the labeling by intermediate subgroup IRs is unambiguous. The cases of repeated IRs are investigated and SALCs for 24-fold coordination in $O_h$ symmetry with account for additional quantum numbers are constructed. The physical significance of this additional quantum numbers is discussed.

Theory

The atoms in a symmetric nanostructures are located at points with local symmetry $H$, which coincides with the full group $G$, or with any of its subgroup, including the subgroup consisting of an identity element only. The whole symmetry group is decomposed into the left cosets with respect to $H$:

$$G = \sum_{i=1}^{n} s_i H, \quad i = 1,\ldots |G|/|H| \qquad (1)$$

Where the module sign stands for the number of elements in a group.

Let $D_k$ be $n_k$ dimensional IR of subgroup $H$ with a basis set $\varphi_\mu$ ($\mu=1,\ldots n_k$). The left coset representatives in (1) move the initial atom $L$ to all equivalent positions $L_i$, which form the orbit of atom $L$. Symmetrical nanoparticle may consist of several orbits, characterizing by different local symmetry subgroups. The number of atoms, forming the orbit of the atom L is given by following relation [3]:

$$N_L = \frac{|G|}{|H|} \qquad (2)$$

According to the theorem on induced representation [8] a set of functions $s_i\varphi_\mu$ ($i=1\ldots N_L$, $\mu=1,\ldots n_k$,) is invariant in the group $G$ and forms the basis of induced representation $D_k \uparrow G$, given by the formula:

$$(D_k \uparrow G)(g)_{i\mu,j\nu} = D_k\left(s_i^{-1} g s_j\right)_{\mu\nu} \delta(s_i^{-1} g s_j) \qquad (3)$$

Where $\delta(s_i^{-1} g s_j) = \begin{cases} 1, & \text{if } s_i^{-1} g s_j \in G \\ 0, & \text{if } s_i^{-1} g s_j \notin G \end{cases}$

In a general case the induced representation (3) is reducible in $G$, and is decomposed into the sum of its IRs $\Gamma_q$:

$$D_k \uparrow G = \sum_q f_q \Gamma_q \qquad (4)$$

Where $f_q$ stand for frequencies of appearance of IRs $\Gamma_q$. Note, that the same basis set is obtained by the projection operator technique [15]. For the decomposition of induced representation (3) one can apply Frobenius reciprocity theorem [7-9], whereby the frequency $f_q$ of IR $\Gamma_q$ of the whole group in the decomposition of induced representations $D_k \uparrow G$ equals to the frequency of appearance of IR $D_k$ in the decomposition of IR $\Gamma_q$ subduced $H$:

$$f(\Gamma_q | D_k \uparrow G) = f(D_k | \Gamma_q \downarrow H) \qquad (5)$$

Thus making use of Frobenius reciprocity theorem the analysis of the SALCs can be done on the basis of orthogonality of characters of $D_k$ and $\Gamma_q$ on the local subgroup $H$ only.

In the case of $I_h$ symmetry group there are orbits, consisting of 12, 20, 30, 60 and 120 atoms, defined by local subgroups: $C_{5v}$, $C_{3v}$, $C_{2v}$, $C_s$ and a subgroup consisting of identity element only. Starting from $s$-orbitals of any atomic center, belonging to totally symmetric IR of the local subgroup and making use of Frobenius reciprocity theorem one easily obtains the IRs for SALCs presented in Table 1. It is seen from Table 1, that in the cases of 30-fold and 60-fold coordinations some IRs appear more then once. In the case of atoms, located in nonsymmetrical points the induced representation is the regular representation and the number of appearance of each IR equals to its dimension [16]. In $O_h$ symmetry repeating IRs for σ-orbitals appear in the case of 24-fold coordination, i.e. when atoms are on the symmetry planes beyond the symmetry axis [4]. Hence it follows that in the case of high total symmetry and low local symmetry additional quantum numbers are required for classification of basis functions of repeating IRs.

To label repeating representations it is possible to use the transitivity property of induction, which means that the two ways of induction (in our case, constructing SALCs from $H$ directly into $G$ and via an intermediate subgroup $F$, provide the same IRs:

$$D_k \uparrow G \propto (D_k \uparrow F) \uparrow G \qquad (6)$$

After inducing into the intermediate subgroup $F$ we may decompose the result into IRs $B_p$ of $F$:

$$D_k \uparrow F = \sum_p f'_p B_p \qquad (7)$$

When inducing each one of IRs $B_p$ into $G$ we obtain :

$$B_p \uparrow G = \sum_q f''_{pq} \Gamma_q \qquad (8)$$

According to the transitivity of induction (6) the frequencies in (7) and (8) are connected by the relation:

$$f_q = \sum_p f'_p f''_{pq} \qquad (9)$$

If frequencies in two-step induction in (7) and (8) do not exceed unity, every repeating IR in (4) acquires unique additional quantum number -index $p$ of IR of intermediate subgroup.

We consider as an example $I_h$ group and atoms located on the planes of symmetry. It is seen in Table 1 that IRs $T_{2u}$, $G_g$ and $G_u$ appear twice and IR $H_g$ appear thrice. To obtain additional quantum numbers for σ-orbitals we consider intermediate induction from local subgroup to intermediate subgroup $C_{5v}$, which results:

$$A \uparrow C_{5v} = A_1 + E_1 + E_2 \qquad (10)$$

Final step of induction of all three IRs from $C_{5v}$ to $I_h$ results:

$$A_1 \uparrow I_h = A_g + T_{1u} + T_{2u} + H_g \qquad (11)$$

$$E_1 \uparrow I_h = T_{1g} + T_{1u} + H_g + H_u + G_g + G_u \qquad (12)$$

$$E_2 \uparrow I_h = T_{2g} + T_{2u} + H_g + H_u + G_g + G_u \qquad (13)$$

In this case repeating IRs do not appear in both steps of the induction and repeating IRs in SALCs get unambiguous classification according to IRs of intermediate group $C_{5v}$. There are three IRs $H_g$, characterized by three different IRs of intermediate group $C_{5v}$, two IRs $T_{1u}$ are labeled by IRs $A_1$ and $E_1$, two IRs $T_{2u}$ by IRs $A_1$ and $E_2$, also two repeating IRs $G_g$ and two repeating IRs $G_u$ are labeled by IRs $E_1$ and $E_2$. It should be noted, that when considering another induction scheme via $C_{3v}$ subgroup one obtains repeating IRs and the labeling according to IRs of $C_{3v}$ is ambiguous.

In the case of $O_h$ group repeating IRs for σ-orbitals appear if atoms are beyond the symmetry axis [4], i.e. for 24-fold coordination (on symmetry planes) and 48-fold coordination (atoms in nonsymmetrical points).

We consider the case of 24-fold coordination on symmetry planes. There are two types of symmetry planes in $O_h$ group - diagonal symmetry planes and coordinate symmetry planes. If the atoms are on diagonal symmetry planes, one obtains following set of IRs:

$$A \uparrow O_h = A_{1g} + A_{2u} + E_g + E_u + T_{1g} + 2T_{1u} + 2T_{2g} + T_{2u} \qquad (14)$$

Since the reflection in diagonal plane belongs to subgroups group $C_{3v}$ and $C_{4v}$, each one of these groups can be chosen as intermediate group.

If the group $C_{3v}$ has physical significance the induction of IR $A$ of group $C_s$ into $C_{3v}$ results the sum of IRs $A_1$ and $E$.

Further induction of IRs $A_1$ and E into group $O_h$ results:

$$A_1 \uparrow O_h = A_{1g} + A_{2u} + T_{1u} + T_{2g} \tag{15}$$

$$E \uparrow O_h = E_g + E_u + T_{1g} + T_{1u} + T_{2g} + T_{2u} \tag{16}$$

No repeating IRs appear in both steps of induction. In this way we obtain unambiguous classification of repeated SALC with respect to IRs of group $C_{3v}$: $T_{1u}(A_1)$, $T_{1u}(E)$, $T_{2g}(A_1)$ and $T_{2g}(E)$.

If the group $C_{4v}$ is chosen as intermediate subgroup, the induction from $C_s$ to $C_{4v}$ results IRs $A_1$, $B_2$ and E. Further induction into $O_h$ results:

$$A_1 \uparrow O_h = A_{1g} + E_g + T_{1u} \tag{17}$$

$$B_2 \uparrow O_h = A_{2u} + E_u + T_{2g} \tag{18}$$

$$E \uparrow O_h = T_{1g} + T_{1u} + T_{2g} + T_{2u} \tag{19}$$

Thus if the group $C_{4v}$ is physically significant the classification of repeating IRs according to group $C_{4v}$ is as follows $T_{1u}(A_1)$, $T_{1u}(E)$, $T_{2g}(B_2)$ and $T_{2g}(E)$.

If 24 atoms are on coordinate planes (see Figure 1) one obtains following set of σ-orbitals:

$$A_1 \uparrow O_h = A_{1g} + A_{2g} + E_g + T_{1g} + 2T_{1u} + T_{2g} + 2T_{2u} \tag{20}$$

In this case the local subgroup belongs to $C_{4v}$, which will be used for searching of additional quantum numbers. The induction of A into $C_{4v}$ results in this case $A_1$, $B_1$ and E. Further induction results:

$$A_1 \uparrow O_h = A_{1g} + E_g + T_{1u} \tag{21}$$

$$B_1 \uparrow O_h = A_{2u} + E_g + T_{2u} \tag{22}$$

$$E \uparrow O_h = T_{1g} + T_{1u} + T_{2g} + T_{2u} \tag{23}$$

Thus we have following labeling of repeating IRs $E_g(A_1)$, $E_g(B_1)$, $T_{1u}(A_1)$, $T_{1u}(E)$, $T_{2u}(B_1)$ and $T_{2u}(E)$.

To elucidate the physical meaning of the additional quantum numbers we construct corresponding SALCs for the latter case considered. The orbit consisting of 24 atoms lying on coordinate planes is shown in Figure 1. The atoms are denoted as follow. The numbers 1, 2, 3, 4, 5 and 6 correspond to directions X, Y, Z, -X, -Y, -Z respectively. The atoms on the upper face of the cube $α_3$, $β_3$, $γ_3$, and $δ_3$ correspond to directions x, y, -x, -y from the middle point of the upper face. We choose two rotations around the axis (111), their products on inversion and pure inversion to transform the functions $\alpha_3$, $\beta_3$, $\gamma_3$, and $\delta_3$ to the corresponding functions $\alpha_i$, $\beta_i$, $\gamma_i$, and $\delta_i$ _ on other faces of cube. The SALSs can be constructed making use of generalized Frobenius reciprocity theorem as follows [9]. Firstly we construct linear combinations of *s*- functions corresponding to the IRs of the local subgroup. This procedure to quite straightforward and the result is written as:

$$\varphi_3(A_1) = \frac{1}{2}\left(s_{\alpha_3} + s_{\beta_3} + s_{\gamma_3} + s_{\delta_3}\right) \quad (24)$$

$$\varphi_3(B_1) = \frac{1}{2}\left(s_{\alpha_3} - s_{\beta_3} + s_{\gamma_3} - s_{\delta_3}\right) \quad (25)$$

$$\varphi_{3x}(E) = \frac{1}{\sqrt{2}}\left(s_{\alpha_3} - s_{\gamma_3}\right)$$
$$\varphi_{3y}(E) = \frac{1}{\sqrt{2}}\left(s_{\beta_3} - s_{\gamma_3}\right) \quad (26)$$

Then we find the projections of basis functions (24), (25) and (26) on the rows of IRs of the whole group, whose coefficients form single columns. The projection coefficients of the basis functions of other local subgroups are obtained by transforming these columns by the matrix of left coset representatives, provided the mapping of the basis functions of other local subgroups is done by the same left coset representatives (in our case rotations around axis (111) and inversion) [9]. The results are presented in Table 2.

Discussion

Comparison of the SALCs presented in Table 2 with standard molecular orbitals in six-fold coordination in $O_h$ -symmetry [15] makes possible to envisage the results of present paper. The influence of the AQN (additional quantum number) is seen explicitly in the Table 2 . SALCs with AQN $A_1$ are similar σ-orbitals constructed from s- or $p_z$ - orbitals and SALC with AQN $E$ are similar to π-orbitals constructed from $p_x$- and $p_y$- orbitals in the case of six-fold coordination [15]. This similarity follows from the same symmetry of our basis set $(\varphi_{3x}, \varphi_{3y})$ and basis set $(p_{3x}, p_{3y})$ [15] of the group $C_{4v}$. On the other hand, SALCs $A_{2u}(B_1)$, $E_g(B_1)$ and $T_{2g}(B_1)$ do not appear in the set of orbitals [15].

In order to envisage the significance of AQNs let us suppose that additional atoms $L_i$ (*i*=1,2...6) are located in coordinate directions (see Figure 1). In this case additional quantum numbers will correspond to hybridizations with orbitals of this atoms. SALCs with AQN $A_1$ interact with s- $p_z$- and $d_{z^2}$ - orbitals. SALCS with AQN $E$ interact with $(p_x, p_y)$ and $(d_{xz}, d_{yz})$ sets of orbitals. SALC with AQN $B_1$ interact with $d_{x^2-y^2}$ orbitals.

Index of the IR of the whole group is a good quantum number in the case when the energy levels of complex are defined by the interaction with central atom [15]. In large nanoparticles the central atom may be absent or the distance between it and the atoms of any shell may be large. In this case the interactions with nearest neighbors in nanonaprticle shell define the energy. Thus for large nanoparticles the energy of electronic orbital is defined with the larger extent by the AQN i.e. by the labels of IR local subgroup.

Conclusion

It was shown that in symmetrical nanoparticles additional quantum numbers are required to label repeating IRs in SALCs constructed from atomic orbitals. The use of intermediate group in the induction makes possible to obtain additional quantum numbers - the labels of IR of intermediate group. In the case when the central atom is absent additional quantum numbers may be more significant for the energy, then the labels of IR of the whole group, since the former define the interaction with nearest neighbors.

References

1. Moses M.J., Fettinger J.C., Eichhorn B.W. Science 2003, 300(5620), 778-781.

2. Kartunen A., Linnolahti M., Pakkanen T.A., Pyykko P. Chem. Comm. 2008, 465, 465-467.

3. Curtis C.W., Reiner I. Representation Theory of Finite Groups and Associative Algebras, Wiley, 1969.

4. Yarzhemskii V.G., Murav'ev E.N. Russian J. Inorg.Chem. 2009, 54, 1341-1344.

5. Racah G. Group theory and spectroscopy. Springer tracts in modern physics 1965, 37, 28-28.

6. Kustov E.F., Yarzhemsky V.G., Nefedov V.I. Int. J. Theoretical Physics 2006, 45(12), 2343-2356.

7. Lederman W. Introduction to Group Characters, Cambridge Univ. Press. 1987.

8. Altman S. L. Induced Representations for Crystals and Molecules. N. Y. Academic. 1977

9. Yarzhemsky V.G., Murav'ev E.N. Doklady Academii Nauk 1984, **278**, 945-948.

10. Yarzhemsky V.G. Few-Body Systems 1997, 22, 27-36.

11. Yarzhemsky V.G., Nefedov V.I. Int. J. Quant. Chem. 2004, 100, 519-527.

12. Chan K.S. and Newman D.J., J. Phys. A: Math. Gen. 1984, 17, 253-265.

13. Newman D.J. J. Phys. A: Math. Gen. 1983, 16, 2375-2387.

14. Newman D.J. and Ng B. Molecular Physics. 1987, 61, 1443.

15. Ballhausen CJ, Introduction to ligand field theory, N.Y. McGraw-Hill,1962.

16. Hamermesh M. Group Theory and Its Application to Physical Problems, Massachusetts, Addison-Wesley 1964.

Table 1. Local subgroups $H$, numbers of equivalent atoms $N$ in orbit and IRs for σ-orbitals in $I_h$ symmetry

| H | N | IR |
|---|---|---|
| $C_{5v}$ | 12 | $A_g+T_{1u}+T_{2u}+H_g$ |
| $C_{3v}$ | 20 | $A_g+T_{1u}+T_{2u}+G_g+G_u+H_g$ |
| $C_{2v}$ | 30 | $A_g+T_{1u}+T_{2u}+G_g+G_u+2H_g+H_u$ |
| $C_s$ | 60 | $A_g+T_{1g}+T_{2g}+2T_{1u}+2T_{2u}+2G_g+2G_u+3H_g+2H_u$ |

Table 2. SALC for the orbit of 24 atoms on coordinate planes in $O_h$ symmetry. Notations correspond to Figure 1 and formulas (24)-(26). IR – irreducible representation of $O_h$ group. AQN - additional quantum number - the label of IR of group $C_{4v}$.

| IR | AQN | SALC |
|---|---|---|
| $A_{1g}$ | $A_1$ | $\frac{1}{\sqrt{6}}[\varphi_1(A_1)+\varphi_2(A_1)+\varphi_3(A_1)+\varphi_4(A_1)+\varphi_5(A_1)+\varphi_6(A_1)]$ |
| $A_{2u}$ | $B_1$ | $\frac{1}{\sqrt{6}}[\varphi_1(B_1)+\varphi_2(B_1)+\varphi_3(B_1)-\varphi_4(B_1)-\varphi_5(B_1)-\varphi_6(B_1)]$ |
| $E_g$ | $A_1$ | $\frac{1}{2\sqrt{3}}[2\varphi_3(A_1)+2\varphi_6(A_1)-\varphi_1(A_1)-\varphi_2(A_1)-\varphi_4(A_1)-\varphi_5(A_1)]$ |
| | | $\frac{1}{2}[-\varphi_1(A_1)+\varphi_2(A_1)-\varphi_4(A_1)+\varphi_5(A_1)]$ |
| $E_g$ | $B_1$ | $\frac{1}{2}[-\varphi_1(B_1)+\varphi_2(B_1)-\varphi_4(B_1)+\varphi_5(B_1)]$ |
| | | $\frac{1}{2\sqrt{3}}[2\varphi_3(B_1)+2\varphi_6(B_1)-\varphi_1(B_1)-\varphi_2(B_1)-\varphi_4(B_1)-\varphi_5(B_1)]$ |
| $T_{1g}$ | $E$ | $\frac{1}{2}[\varphi_{3y}(E)-\varphi_{2x}(E)+\varphi_{6y}(E)-\varphi_{5y}(E)]$ |
| | | $\frac{1}{2}[-\varphi_{3x}(E)+\varphi_{1y}(E)-\varphi_{6x}(E)+\varphi_{4y}(E)]$ |
| | | $\frac{1}{2}[-\varphi_{1x}(E)+\varphi_{2y}(E)-\varphi_{4x}(E)+\varphi_{5y}(E)]$ |
| $T_{1u}$ | $A_1$ | $\frac{1}{\sqrt{2}}[\varphi_1(A)-\varphi_4(A)]$ |
| | | $\frac{1}{\sqrt{2}}[\varphi_2(A)-\varphi_5(A)]$ |
| | | $\frac{1}{\sqrt{2}}[\varphi_3(A) - \varphi_6(A)]$ |
| $T_{1u}$ | $E$ | $\frac{1}{2}[\varphi_{3x}(E) + \varphi_{2y}(E) - \varphi_{6x}(E) - \varphi_{5y}(E)]$ |
| | | $\frac{1}{2}[\varphi_{1x}(E) + \varphi_{3y}(E) - \varphi_{4x}(E) + \varphi_{6y}(E)]$ |
| | | $\frac{1}{2}[\varphi_{2x}(E) + \varphi_{1y}(E) - \varphi_{5x}(E) - \varphi_{4y}(E)]$ |
| $T_{2g}$ | $B_1$ | $\frac{1}{\sqrt{2}}[\varphi_1(B) + \varphi_4(B)]$ |
| | | $\frac{1}{\sqrt{2}}[\varphi_2(B) + \varphi_5(B)]$ |
| | | $\frac{1}{\sqrt{2}}[\varphi_3(B) + \varphi_6(B)]$ |
| $T_{2g}$ | $E$ | $\frac{1}{2}[\varphi_{2x}(E) + \varphi_{3y}(E) + \varphi_{5x}(E) + \varphi_{6y}(E)]$ |
| | | $\frac{1}{2}[\varphi_{3x}(E) + \varphi_{1y}(E) + \varphi_{6x}(E) + \varphi_{4y}(E)]$ |
| | | $\frac{1}{2}[\varphi_{1x}(E) + \varphi_{2y}(E) + \varphi_{4x}(E) + \varphi_{5y}(E)]$ |
| $T_{2u}$ | $E$ | $\frac{1}{2}[\varphi_{3x}(E) - \varphi_{2y}(E) - \varphi_{6x}(E) + \varphi_{5y}(E)]$ |
| | | $\frac{1}{2}[\varphi_{1x}(E) - \varphi_{3y}(E) - \varphi_{4x}(E) + \varphi_{6y}(E)]$ |
| | | $\frac{1}{2}[\varphi_{2x}(E) - \varphi_{1y}(E) - \varphi_{5x}(E) + \varphi_{4y}(E)]$ |

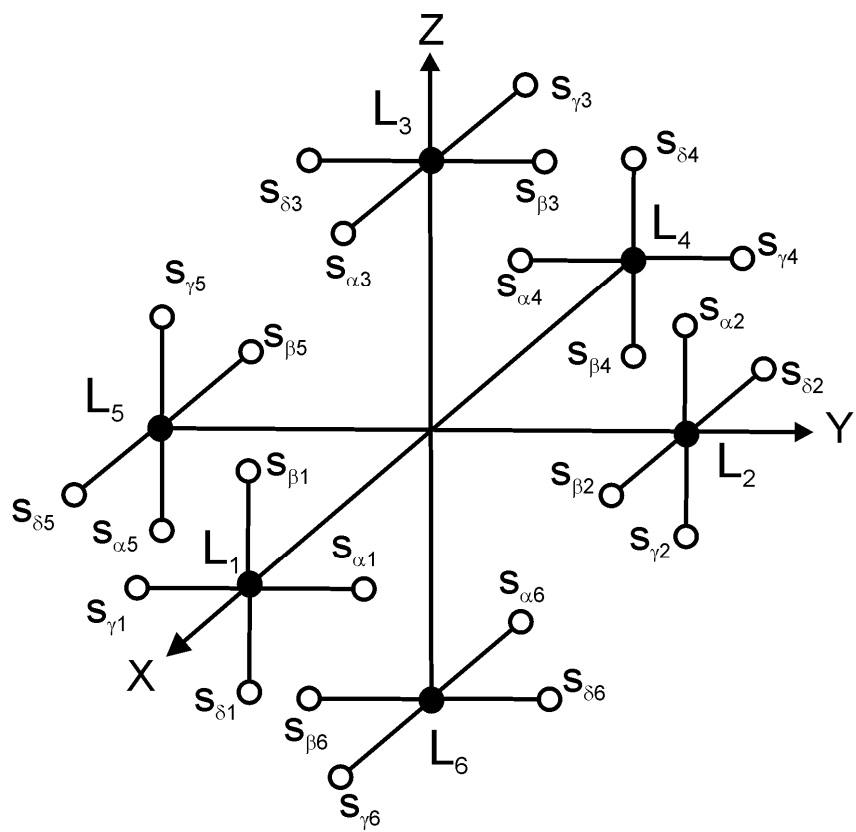

Figure 1. Coordinate system and two orbits of 6 and 24 atoms on coordinate planes in $O_h$ symmetry.